# Field-driven collapsing dynamics of skyrmions in magnetic multilayers

R. Tomasello[1,*,#], Z. Wang[2,3,*], E. Raimondo[4], S. Je[5], M. Im[5], M. Carpentieri[1], W. Jiang[2,3,#], G. Finocchio[4,#]

[1]Department of Electrical and Information Engineering, Politecnico di Bari, Bari 70125, Italy

[2]State Key Laboratory of Low-Dimensional Quantum Physics and Department of Physics, Tsinghua University, Beijing 100084, China

[3]Frontier Science Center for Quantum Information, Tsinghua University, Beijing 100084, China

[4]Department of Mathematical and Computer Sciences, Physical Sciences and Earth Sciences, University of Messina, I-98166, Messina, Italy

[5]Center for X-ray Optics, Lawrence Berkeley National Laboratory, Berkeley, CA 94720, USA

[*]These authors contributed equally.

**Abstract**

Magnetic skyrmions are fascinating topological particle-like textures promoted by a trade-off among interfacial properties (perpendicular anisotropy and Dzyaloshinskii-Moriya interaction (DMI)) and dipolar interactions. Depending on the dominant interaction, complex spin textures, including pure Néel and hybrid skyrmions have been observed in multilayer heterostructures. A quantification of these different spin textures typically requires a depth-reoslved magnetic imaging or scattering techniques. In the present work, we will show qualitatively different collapsing dynamics for pure Néel and hybrid skyrmions induced by a perpendicular magnetic field in two representative systems, [Pt/Co/Ir]$_{15}$ and [Ta/CoFeB/MgO]$_{15}$ multilayers. Skyrmions in the former stack undergo two morphological transitions, upon reversing the perpendicular field direction. Skyrmions in [Ta/CoFeB/MgO]$_{15}$ multilayers exhibit a continuous transition, which is mainly linked to a reversible change of the skyrmion size. A full micromagnetic phase diagram is presented to identify these two collapsing mechanisms as a function of material parameters. Since the two distinct collapsing dynamics rely on the detailed layer-dependent spin structures of skyrmions, they could be used as potential fingerprints for identifying the skyrmion type in magnetic multilayers. Our work suggests the employment of pure and hybrid skyrmions for specific applications, due to the strong correlation between the skyrmion dynamics and 3-dimentional spin profiles.

## 1. Introduction

Magnetic skyrmions have been receiving growing attentions in the last decades. Since the first theoretical predictions on the fundamental and promising properties of these "topologically protected" solitons[1–4], great efforts have driven the development of materials for the experimental stabilization of skyrmions. The first experimental observations regarded $B_{20}$ compounds, where the bulk Dzyaloshinskii-Moriya interaction (DMI) promotes the formation of Bloch skyrmions[5–10]. Later on, increasing interests have been devoted to thin films and heterostructures where ferromagnetic layers (FM) are coupled with materials having strong spin-orbit coupling including heavy metals (HM)[11]. These latter systems allow for the stabilization of skyrmions at room temperature and above, which is a fundamental requirement for practical applications[12]. For example, skyrmions were observed in a variety of thin film materials, including HM/single-layer FM/oxide[13–15], $HM_1$/FM/$HM_2$ multilayer heterostructures[16–24], HM/ferrimagnet/oxide multilayers[25,26] synthetic antiferromagnets[27–29], as well as hybrid systems[30,31]. The use of $HM_1$/FM/$HM_2$ multilayer heterostructures, with the right choice of $HM_1$ and $HM_2$, for stabilizing skyrmion is favorable over other solutions for two reasons[16]: (*i*) the use of asymmetric interfaces, by combining HMs having an opposite sign of interfacial DMI, enhances the effective DMI, thus allowing the stabilization of sub 100 nm skyrmion; (*ii*) the skyrmion magnetic volume increases since the skyrmion extends along the thickness direction through the whole multilayer, as a result of the magnetostatic coupling. This allows the skyrmion to be less sensitive to the thermal fluctuations. The static skyrmion profile and chirality in $HM_1$/FM/$HM_2$ multilayers result from the competition between magnetostatic and interfacial DMI fields[23,32]. At zero or low interfacial DMI (IDMI) values, the magnetostatic field is dominant and skyrmions exhibit a thickness-dependent spin chirality along the thickness direction of the multilayer, which are known as hybrid skyrmions. When the IDMI is high, homogenous Néel skyrmions throughout the thickness – known as pure Néel skyrmions[32,33] – can be achieved.

The different chirality of skyrmions can strongly influence their current-driven dynamics[32,34]. Therefore, understandings of three-dimensional profile of skyrmions in different materials systems are crucial for spintronic applications. The thickness-dependent chirality can be directly observed in domain walls via the circular dichroism x-ray resonant magnetic scattering[23,32], those results are then used to identify the type of skyrmions, i.e., hybrid or pure Néel. Recently, an indirect procedure based on First Order Reversal Curve (FORC) technique has been proposed, which suggests systems with pure Néel skyrmion and hybrid skyrmion exhibits different characteristics in the FORC diagrams[33]. Here, we propose a strategy to identify the type of skyrmions stabilized in a magnetic multilayer, i.e., pure Néel or hybrid skyrmions, by performing imaging of the evolution of the spin textures as a function of out-of-plane magnetic fields. Two typical multilayers were characterized experimentally: [Pt/Co/Ir]$_{15}$ with a stronger IDMI and [Ta/CoFeB/MgO]$_{15}$ with a lower IDMI, respectively. The former shows a morphological transition from the skyrmion state to the labyrinth domains. By contrast, a reversible skyrmion state that expands or shrinks according to the field is observed in the latter system. We further perform micromagnetic simulations to explore the thickness-dependent chirality of skyrmions, which confirm the presence of two different types of skyrmions. A detailed understanding of the collapsing dynamics is obtained through calculating the complete phase diagram that summarizes the dynamics on material specific parameters. Our results can be used as a procedure for characterizing the spin profile of skyrmions in multilayers, which could be useful for future skyrmionic applications, since the dynamics of skyrmions, such as the skyrmion Hall angle, are deeply affected by their static three-dimensional profiles.

## 2. Results and Discussion

*Hysteresis measurements.* Samples #1 [Pt(1.5 nm)/Co(1 nm)/Ir(1.5 nm)]$_{15}$ and #2 [Ta(3 nm)/Co$_{20}$Fe$_{60}$B$_{20}$(1 nm)/MgO(2 nm)]$_{15}$ multilayers (values in parentheses present the thickness) were considered (see section 4 for the details), as schematics illustrated in Figs. 1(a) and (c), respectively. The magnetic properties of these samples were characterized, results of which are shown in Figs. 1(b) and (d) for the multilayers [Pt/Co/Ir]$_{15}$ and [Ta/Co$_{20}$Fe$_{60}$B$_{20}$/MgO]$_{15}$, respectively. There, one can find the corresponding in-plane and out-of-plane magnetic hysteresis loops, which are consistent with typical skyrmion-hosting multilayers.

A series of the [Pt/Co/Ir]$_{15}$ magnetic images collected at the Co L$_3$ edge (778.5 eV) under perpendicular fields $H_\perp$ is given in Fig. 2(a). At $\mu_0 H_\perp = 141$ mT, we observe the presence of isolated quasi-circular Néel skyrmions. As magnetic fields decrease to zero, these Néel skyrmions are expanded and deformed in shape, i.e., a mixed phase of isolated larger skyrmions and elongated ones are obtained. Upon inverting the direction of magnetic field, this phase is converted into a labyrinthine domain configuration (at $\mu_0 H_\perp = -92.3$ mT). A further increase of magnetic field dissects labyrinthine domains into isolated skyrmions with a revered magnetization of the core of skyrmion.

By contrast, very different collapsing phenomenon has been found in the [Ta/CoFeB/MgO]$_{15}$ multilayer. Its magnetic images were conducted at the Fe L3 edge (708.5 eV) under varied $H_\perp$, as shown in Fig. 2(b). The existence of circular domains at $H_\perp = 47.8$ mT is assumed to be the isolated hybrid skyrmions. Initially, the size of these skyrmions increases with a decreased $H_\perp$ magnitude till 0 mT. As the direction of field is reversed, which then becomes parallel to the skyrmion core direction, the hybrid skyrmions continuously expand. A further increase of $H_\perp$ results in the skyrmions to coalesce and form big band domains, with the core magnetization orienting along the field direction (at $\mu_0 H_\perp = -67.0$ mT).

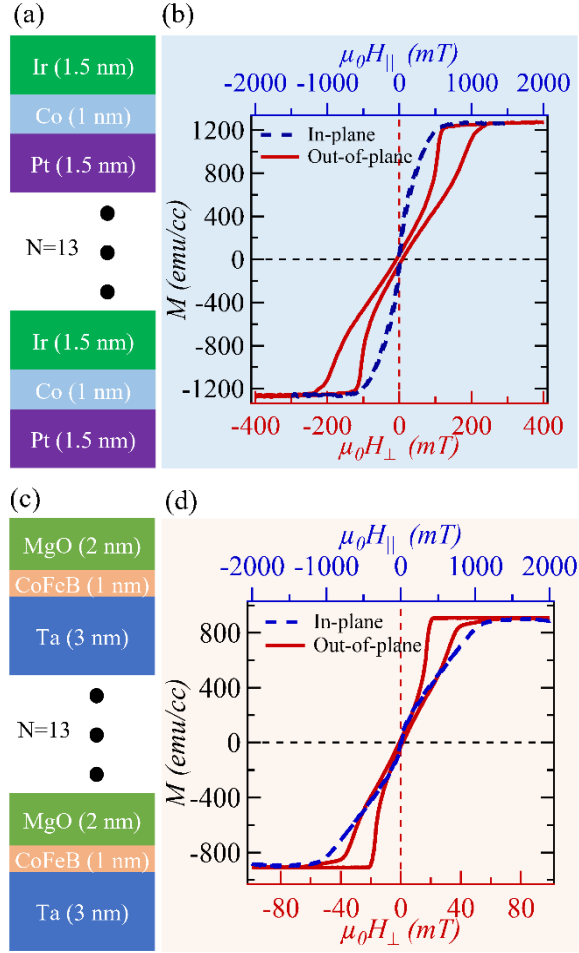

**Figure 1.** (a) Schematic illustration of the [Pt(1.5 nm)/Co(1 nm)/Ir(1.5 nm)]$_{15}$ multilayer heterostructure. (b) The obtained in-plane and out-of-plane magnetic hysteresis loops. (c) Schematic illustration of the [Ta(3 nm)/ Co$_{20}$Fe$_{60}$B$_{20}$ (1 nm)/MgO(1.5 nm)]$_{15}$ multilayer heterostructure. (d) The corresponding in-plane and out-of-plane magnetic hysteresis loops. The $H_\perp$ and $H_\parallel$ represents the applied magnetic fields that are parallel and perpendicular to the film plane, respectively.

It is clear that, the field-driven collapsing dynamics are dramatically different in the [Pt/Co/Ir]$_{15}$ and [Ta/CoFeB/MgO]$_{15}$ multilayers. In particular, the experimental results suggest skyrmions in the former multilayer which go through two phase transitions: (1) from isolated skyrmions to labyrinth domains, and (2) from labyrinth domains to skyrmions with opposite core to the initial ones. In the latter material system, a reversible process from isolated skyrmions to big domains is observed, in which the magnetization orientation is along the direction of the core of the original skyrmions. These results imply their different spin profiles across the thickness direction that will be studied by performing micromagnetic simulations.

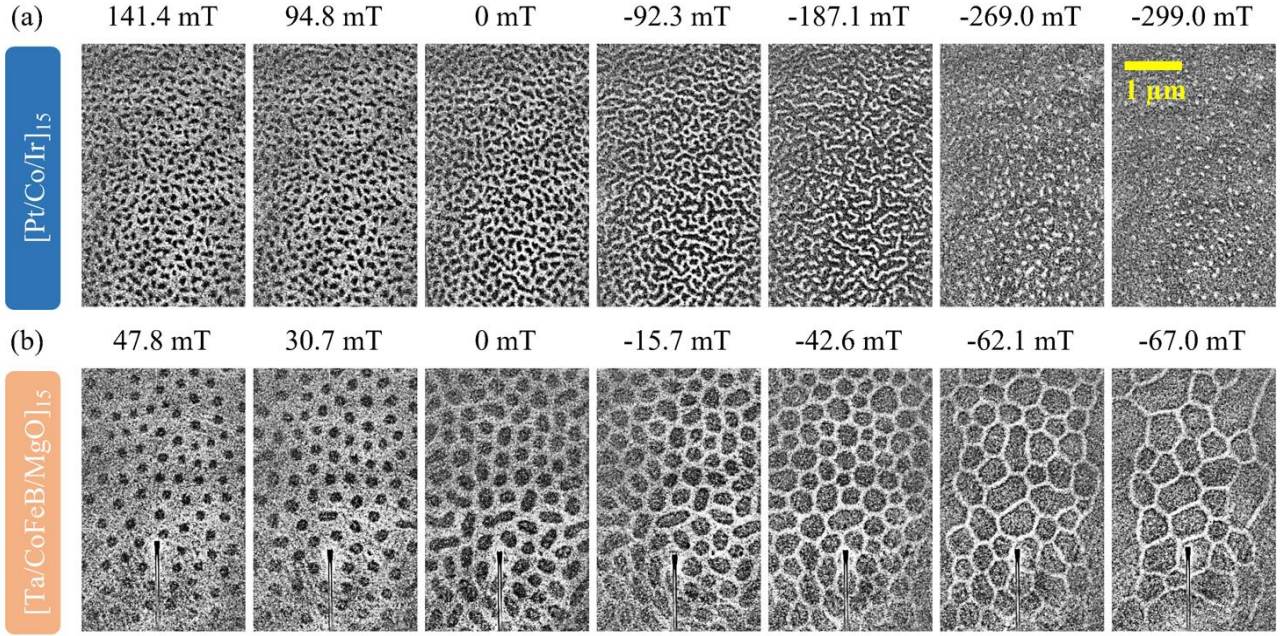

**Figure 2.** Experimental observations of the field-driven collapsing dynamics of skyrmions in (a) [Pt/Co/Ir]$_{15}$ and (b) [Ta/CoFeB/MgO]$_{15}$ multilayers. Note that the white colour corresponds to the local magnetization pointing out of the film ($m > 0$) and vice versa.

*Micromagnetic results – Comparison of pure Néel and hybrid skyrmions.* Below, we perform micromagnetic simulations (see section 4 for the details) and use an analytical theory to reveal the role of the layered-dependent spin chirality in assisting collapsing dynamics of skyrmions in these two material systems.

Figure 3 shows the two equilibrium configurations of skyrmions in these two multilayers. For the sample #1, a Néel skyrmion is observed in all the layers, suggesting the independence of chirality on thickness due to the large contribution from IDMI. In the sample #1, the effect of the magnetostatic field can only be observed in the depth-dependent size of the skyrmion, which is smaller in the external layers, and bigger in the middle layer. On the other hand, for the sample #2 with a smaller IDMI, a hybrid skyrmion is stabilized, where the skyrmion undergoes the transformation along the multilayer thickness from Néel with outward chirality in the bottom layer (layer 1), to Bloch with counterclockwise chirality in layer 4, and to Néel with inward chirality in the top layer (layer 5)[32]. The Bloch skyrmion is not located in the middle layer because of the finite $D = 0.5$ mJ/m$^2$[23,32,33]. Our micromagnetic simulations results confirm the above assumption of the existence of two different types of skyrmions in these two distinct multilayers.

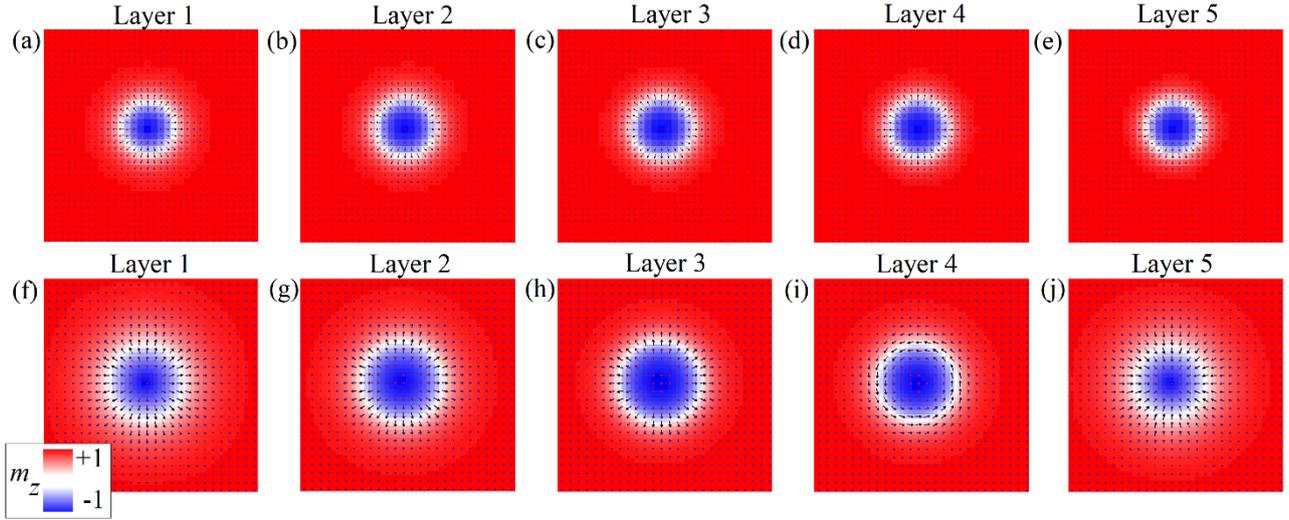

**Figure 3.** Equilibrium configurations of skyrmions as obtained by micromagnetic simulations in the x-y plane. A-e Néel skyrmions in the 5-repetitions [Pt/Co/Ir]$_{15}$ multilayer, for $\mu_0H_\perp$ = 141 mT. e-h Hybrid skyrmions in the 5-repetitions [Ta/CoFeB/MgO]$_{15}$ multilayer, for $\mu_0H_\perp$ = 47.9 mT.

*Micromagnetic results – Field scan as a function of the out-of-plane field.* We start the simulations with an initial state containing 6 pure Néel skyrmions in sample #1 at perpendicular external field $\mu_0H_\perp$ = 141 mT, and 9 hybrid skyrmions in sample #2 at $\mu_0H_\perp$ = 47.8 mT (see Fig. 4). The purpose is to try to reproduce the experimental dynamics in Fig. 2. We gradually reduce the applied field and trace the evolution of the skyrmions. In sample #1, at $\mu_0H_\perp$ = 141 mT, skyrmions are small, isolated and circularly-shaped, which are in agreement with our experimental results and with the ones in high-DMI multilayers[33]. As the field decreases to zero, skyrmions become larger, elongated and deformed in shape. When the field is reversed, such a phase is converted into a labyrinth domain configuration (see $\mu_0H_\perp$ = -92.3 mT). A further increase of field breaks the labyrinth domain into small skyrmions with a reversed core.

In the sample #2, the isolated hybrid skyrmions at $\mu_0H_\perp$ = 47.9 mT increase in size and deformed into an elongated worm-like shape as the field is reduced to zero. When the field is reversed, which then becomes parallel to the skyrmion core direction, the hybrid skyrmions keep expanding until they start to coalesce and form big band domains, with the magnetization orienting along the same direction of the applied field ($\mu_0H_\perp$ = -29.8 mT). All the previous numerical results are in agreement with the experimental observations in Fig. 2.

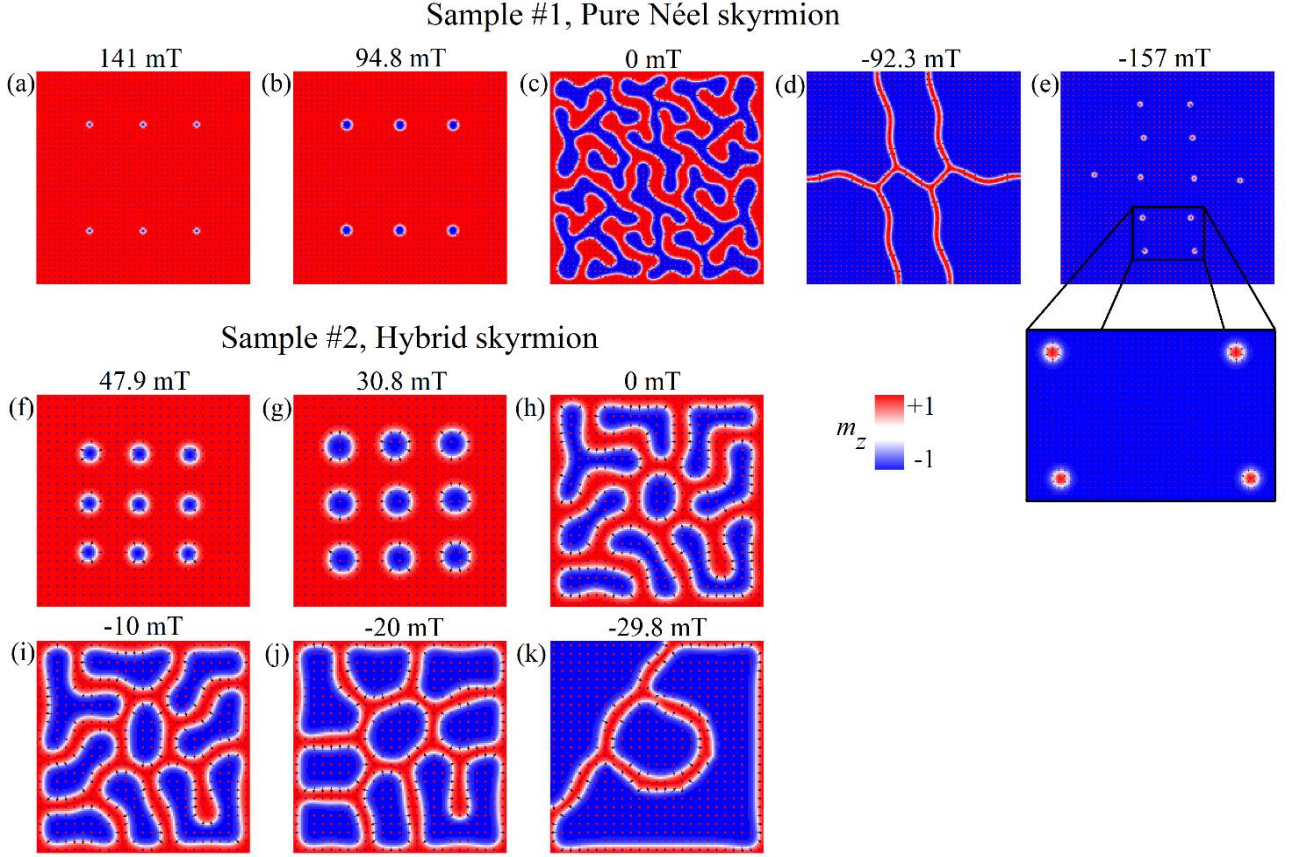

**Figure 4**. Micromagnetic simulation results of the collapsing dynamics in the top layer of the two samples #1 and #2.

*Micromagnetic results – Explanation.* The origin of the two different collapsing dynamics will be studied. In the sample #1, a skyrmion with a Néel chirality in all the layers is stabilized by a large IDMI, therefore its energetic stability can be described by using the same argument as a Néel skyrmion in a single ferromagnetic layer[35]. Figure 5 shows the evolution of energy as a function of the applied field of a single-layer Néel skyrmion and uniform states. This is done by minimizing the energy functional $E[\mathbf{m}] = \int dV \varepsilon(\mathbf{m})$, with the energy density[35]

$$\varepsilon(\mathbf{m}) = A(\nabla \mathbf{m})^2 + \varepsilon_{DMI} + K_u m_z^2 - 0.5 M_s \mathbf{m} \cdot \mathbf{H}_m - M_s \mathbf{m} \cdot \mathbf{H}_{ext}, \qquad (1)$$

$\varepsilon_{DMI} = D[m_z(\nabla \cdot \mathbf{m}) - (\mathbf{m} \cdot \nabla) m_z]$ is the interfacial DMI energy density, $m_z$ is the magnetization $z$-component, $\mathbf{H}_m$ is the magnetostatic field, and $\mathbf{H}_{ext}$ is the external magnetic field. The Néel skyrmion is a metastable state with respect to the uniform state (skyrmion energy is larger than the uniform one) for 4 µT $\leq \mu_0 H_\perp <$ 150 mT, while it becomes a stable state for $0 \leq \mu_0 H_\perp <$ 4 µT (see Fig. 5). These results qualitatively explain the collapsing dynamics in sample #1 (Fig. 2(a), and Fig. 4(a)-(e)). In particular, the pure Néel skyrmions are metastable for $\mu_0 H_\perp \leq$ 141 mT, then become stable in the range of -92.3 mT $< \mu_0 H_\perp <$ 0 mT, giving rise to a phase transition to labyrinth domains, and, eventually become again metastable states for higher fields (the energetic stability is symmetric upon the field reversal).

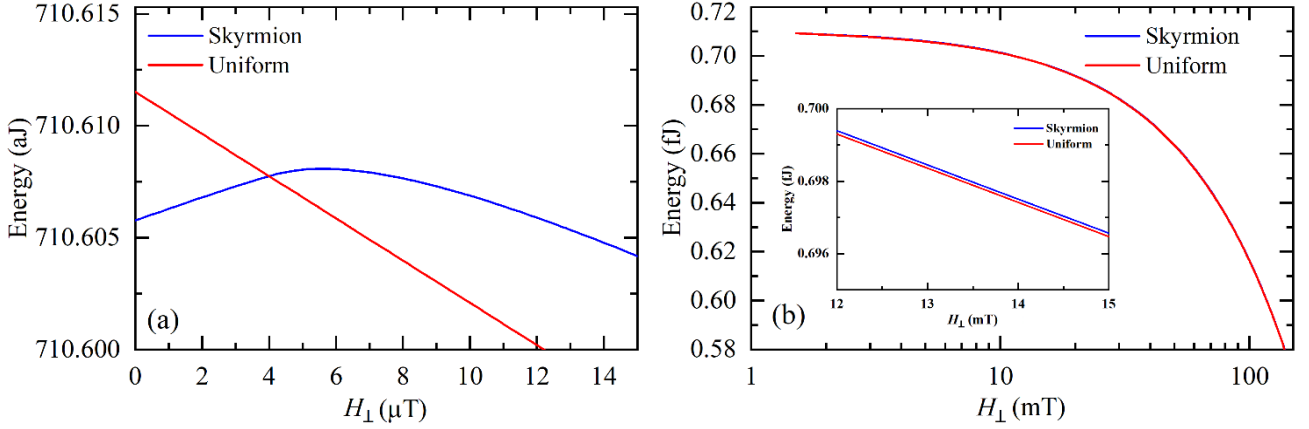

**Figure 5.** Comparison between the skyrmion and uniform energies as a function of the perpendicular field, as computed from the minimization of the energy functional[35] in Eq. (1). (a) Energies in the field range $0\ \mu T \leq \mu_0 H_\perp < 15\ \mu T$, where the transition field of 4 µT is evident. (b) Energies in the field range $0\ mT \leq \mu_0 H_\perp \leq 150\ mT$ in a logarithmic scale. Inset in (b): magnified section of the energies in the field range $12\ mT \leq \mu_0 H_\perp \leq 15\ mT$ to show that the skyrmion is a metastable state.

The effect of external fields on the hybrid skyrmions in sample #2 can be understood from early results on magnetic bubbles in bulk materials[36–40]. In this scenario, we can identify three cases according to the value of the quality factor $Q = 2K_u/\mu_0 M_s^2$:

1. $Q < 1$, the perpendicular anisotropy is weaker than the magnetostatic field, hence the easy-axis is in-plane and vortices are the stable solution in infinite film[41]. However, if the sample dimensions are reduced below a threshold value and the sample thickness is increased, magnetic bubbles with a large domain wall can be still stabilized (the confinement effect is crucial).

2. $1 < Q < 2$, the perpendicular anisotropy is comparable or larger than the magnetostatic field. The easy-axis is out-of-plane and magnetic bubbles are the energetically-favored solution for large sample dimensions, even at zero external field[36].

3. $Q > 2$, the perpendicular anisotropy is much larger than the magnetostatic field. The easy-axis is out-of-plane and magnetic bubbles are observed at zero field, as a result of the confinement effect in sufficiently small samples[38]. Whereas, for the larger dots, random labyrinth domains are obtained at zero field.

In sample #2, the quality factor is 1 or slightly larger, therefore the discussion of the point 2 supports our results. We wish to underline that the energy landscape for stabilizing hybrid skyrmions, and magnetic bubbles is very similar, i.e., they are both energetically-favored solutions of the magnetostatic energy minimization. Therefore, the response of a magnetic bubble to external fields is expected to be qualitatively similar to that one of a hybrid skyrmion in a low/zero IDMI multilayer. Indeed, Fig. 6 shows our simulations for a magnetic bubble of a bulk materials with a perpendicular magnetic anisotropy and zero IDMI, which has the same parameters and geometrical dimensions as the previous multilayer sample #2. The number of magnetic bubbles (Bloch skyrmions) is also the same. We observe a qualitatively similar field response, where magnetic bubbles are initially stabilized by a positive field ($\mu_0 H_\perp$=200 mT). As the field decreases, magnetic bubbles transform into worm-like shapes. After reversing the direction of field, big band domains appear with the core magnetization parallel to the negative field ($\mu_0 H_\perp$=-150 mT).

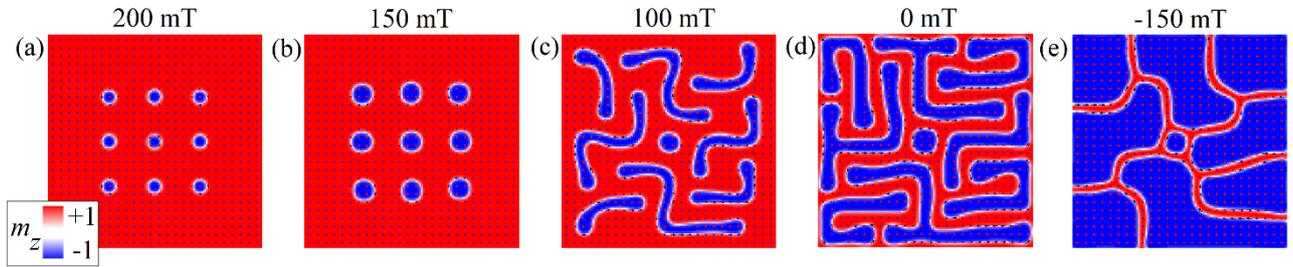

**Figure 6.** Micromagnetic simulations results of the collapsing dynamics of magnetic bubbles in a perpendicular bulk material.

However, we wish to point out that two key differences exist between bulk samples and multilayers hosting hybrid skyrmions, which do not affect our conclusions: (i) in multilayers, each ferromagnetic layer is coupled only through dipolar interactions, and (ii) the IDMI can be only introduced in multilayers due to the presence of asymmetric interfaces where the ferromagnetic layers are thin (typically with a thickness smaller than 1 nm).

To get a deeper understanding of the collapsing dynamics of skyrmions in magnetic multilayers, we perform massive micromagnetic simulations as a function of the material parameters, which is summarized in Fig. 7, where a collapsing dynamics diagram in the $Q - d$ ($d = 2A/D$) space is illustrated[35]. We identify 5 main regions:

1. for $Q < 1$ and $0 \leq d \leq 0.55$ (blue region): the initial state contains hybrid skyrmions (HS) after relaxation and they undergo a collapsing dynamics similar to the sample #1 (collapsing dynamics type 1, whichlead to worm-like domains as the field is reduced to zero, and to a mixed phase as the field is reversed and equal to -100 mT (see Fig. 8). In particular, the mixed phase includes hybrid skyrmions and magnetic bubbles. This is because the IDMI is not sufficiently large to fix the skyrmion helicity throughout the layers.

2. For $1 \leq Q \leq 1.25$ and $0 \leq d \leq 0.5$ (red region): the initial state contains hybrid skyrmions after relaxation. The dynamics type 2 are similar to the results of the sample #2, meaning conversion of the hybrid skyrmions into big band domains as the field decreases and reverses its sign. Indeed, the experimental magnetic parameters of sample #2 fall in this region, as indicated by the black star.

3. For $1.25 < Q \leq 1.4$ and $0.2 \leq d \leq 0.4$ (red region): the initial state contains pure Néel skyrmions (PNS) after relaxation. However, the collapsing dynamics is similar to the sample #2 (collapsing dynamics type 2) and big band domains are obtained in the end.

4. For $1.25 < Q \leq 1.4$ and $0 \leq d \leq 0.3$ (yellow region): the equilibrium configuration is the uniform out-of-plane state because the perpendicular magnetic anisotropy is too high to allow the existence of skyrmions.

5. For $1 \leq Q \leq 1.4$ and $0.5 < d \leq 1$ (green region): the initial state contains pure Néel skyrmions after relaxation. The dynamics type 1 are similar to the results of the sample #1, meaning that skyrmions undergo two phase transitions to labyrinth domains and Néel skyrmions with opposite polarity. Indeed, the experimental magnetic parameters of sample #1 fall in this region, as indicated by the black star.

Based on the full phase diagram, we can conclude that when $1 < Q < 1.25$ for any given choice of $d$, and for $0.6 \leq d \leq 1$ for any given choice of $Q$, it is possible to establish a direct relation between the type of skyrmions and the collapsing dynamics, which are exactly the cases of our experimental observations.

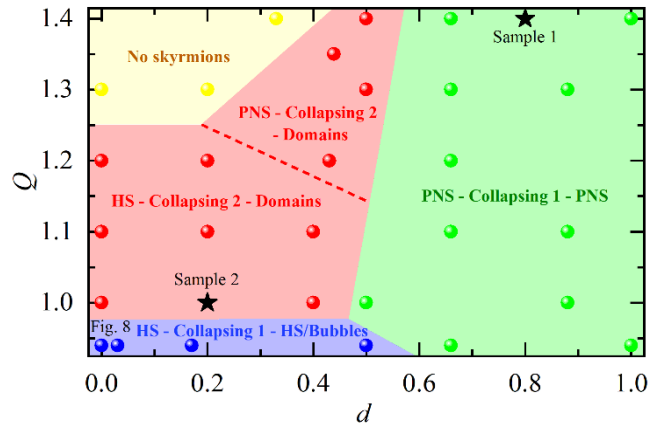

**Figure 7.** A phase diagram that summarizes the collapsing dynamics in the $Q - d$ space, which is obtained by micromagnetic simulations. Different colors are utilized to represent different states (pure Néel skyrmions – PNS –, or hybrid skyrmions - HS) and the associated type of collapsing dynamics (type 1 – similar to sample #1 –, or type 2 – similar to sample #2). The scattered symbols refer to the $Q$ and $d$ values used in the simulations. The black stars indicated the values of $Q$ and $d$ corresponding

to the experimental samples #1 and #2, respectively. An example of collapsing dynamics concerning the blue region is illustrated in Fig. 8.

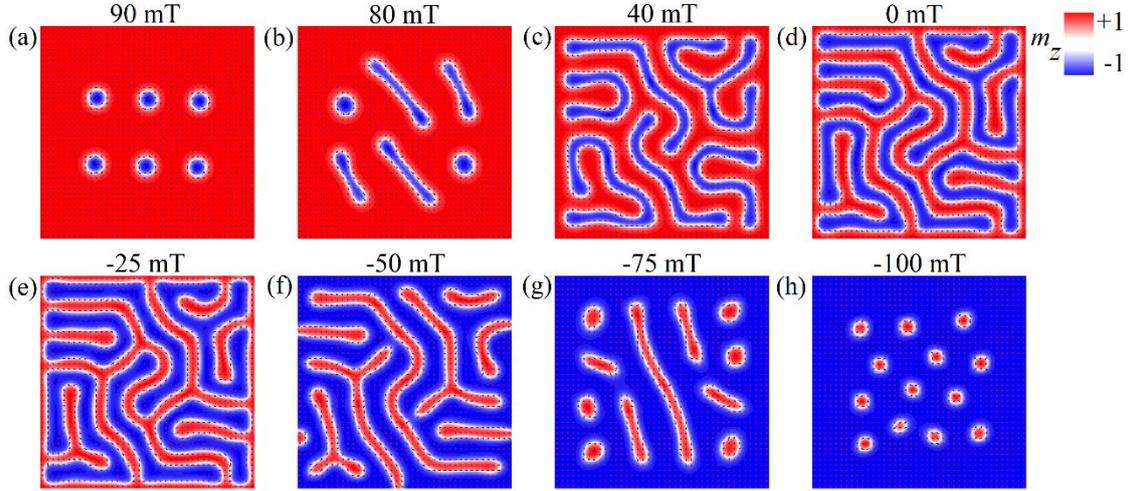

**Figure 8.** Micromagnetic simulations of the collapsing dynamics type 1 of hybrid skyrmions to a mixed phase including hybrid skyrmions and bubbles for $Q = 0.94$ and $d = 0$.

## 3. Summary and Conclusions

In summary, we have studied both experimentally and numerically the collapsing dynamics of magnetic skyrmions in two distinct experimental multilayer heterostructures, where pure Néel and hybrid skyrmions exist. Both experimental observations and numerical results show two different collapsing behaviours which we ascribed to the competition between different energy terms. In IDMI-dominated system - [Pt/Co/Ir]$_{15}$ multilayer heterostructure, such collapsing dynamics are consequences of the change of the energetic stability of pure Néel skyrmions due to the change of the external field. Skyrmions are metastable at $\mu_0 H_\perp = 141$ mT, then become stable or quasi-stable in the range 0 mT $< \mu_0 H_\perp <$ |92.3| mT, giving rise to a morphological transition to labyrinthine domains, and finally become metastable states with opposite cores for higher fields. On the contrary, in magnetostatic-dominated system-[Ta/CoFeB/MgO]$_{15}$ multilayer heterostructure, hybrid skyrmions are characterized by reversible processes where the skyrmion cores expand following an increase of external fields, and merge together to form big band domains. Our work establishes an easily-accessible technique to identify the type of skyrmion in a magnetic multilayer. Our results can be used as preliminary tests preceding the use of skyrmions in specific applications, since the skyrmion dynamics strongly depend on the static equilibrium three-dimensional profile in magnetic multilayers.

## 4. Experimental Section

*Samples description and characterization.* Samples #1 [Pt(1.5 nm)/Co(1 nm)/Ir(1.5 nm)]$_{15}$ and #2 [Ta(3 nm)/Co$_{20}$Fe$_{60}$B$_{20}$(1 nm)/MgO(2 nm)]$_{15}$ multilayers (values in parentheses present the thickness) were deposited onto thermally oxidized silicon substrates using an ultra-high vacuum magnetron sputtering system, as schematics illustrated in Figs. 1(a) and (c), respectively. Both multilayers are known for hosting the nanoscale skyrmions at room temperature[12,16,17], as a result of IDMI. The magnetic properties of these samples were characterized via using a superconductor quantum interference device (SQUID) magnetometer (MPMS, Quantum Design). The saturation magnetization and anisotropy field are estimated to be $M_s$ = 1200 emu/cc and $\mu_0 H_k$ = 500 mT for the [Pt/Co/Ir]$_{15}$ multilayer, and $M_s$ = 900 emu/cc and $\mu_0 H_k$ = 1000 mT for the [Ta/Co$_{20}$Fe$_{60}$B$_{20}$/MgO]$_{15}$ multilayer, respectively. These multilayers were also deposited on the 100 nm-thick Si$_3$N$_4$ membranes (Clean SiN, Suzhou) for imaging characterization by using the full-field soft x-ray transmission microscopy with a spatial resolution approaching 20 nm, which is performed at the beamline 6.1.2, Advanced Light Source, Lawrence Berkeley National Laboratory. This imaging approach allows us to capture the collapsing dynamics of Néel- or hybrid-type skyrmions that are driven by out-of-plane magnetic fields.

*Micromagnetic simulations.* Micromagnetic computations were performed by means of a state-of-the-art micromagnetic solver, PETASPIN, which numerically integrates the Landau-Lifshitz-Gilbert (LLG) equation by applying the Adams-Bashforth time solver scheme[42]:

$$\frac{d\mathbf{m}}{d\tau} = -(\mathbf{m} \times \mathbf{h}_{\text{eff}}) + \alpha_G \left(\mathbf{m} \times \frac{d\mathbf{m}}{d\tau}\right), \qquad (2)$$

where $\mathbf{m} = \mathbf{M}/M_s$ is the normalized magnetization vector, $\alpha_G$ is the Gilbert damping, and $\tau = \gamma_0 M_s t$ is the dimensionless time, with $\gamma_0$ being the gyromagnetic ratio and $M_s$ the saturation magnetization. $\mathbf{h}_{\text{eff}}$ is the normalized effective magnetic field, which includes the exchange, IDMI, uniaxial anisotropy, and external perpendicular magnetic fields, as well as the magnetostatic field computed by calculating the magnetostatic energy of the whole system.

We study two set of samples that were signified by very different material specific parameters, #1 and #2, characterized by 5 repetitions of ferromagnet layer of thickness 1 nm. For the sample #1 hosting a pure Néel skyrmion, we used typical parameters values of [Pt/Co/Ir]$_{15}$ multilayers[16]: saturation magnetization $M_s$ = 1.2 MA/m, $K_u$ = 1.29 MJ/m$^3$, exchange constant $A$ = 10 pJ/m, IDMI parameter $D$ = 2.5 mJ/m$^2$, and a discretization cell size of 2.5 × 2.5 × 1 nm$^3$. Whereas, for the sample #2 hosting a hybrid skyrmion, we used typical parameters values of [Ta/CoFeB/MgO]$_{15}$ multilayers[23]: $M_s$ = 0.91 MA/m, $K_u$ = 0.52 MJ/m$^3$, exchange constant $A$ = 10 pJ/m, IDMI parameter $D = 0.5$ mJ/m$^2$, and a discretization cell size of $4 \times 4 \times 1$ nm$^3$. The ferromagnetic layers are coupled to one another by means of only the magnetostatic field (exchange-decoupled layers), and are separated by a 3 nm and 4 nm thick non-magnetic layer for sample #1 and #2, respectively.

**Acknowledgements**

The research has been supported by the project PRIN 2020LWPKH7 funded by the Italian Ministry of Research, and by the project number 101070287 — SWAN-on-chip — HORIZON-CL4-2021-DIGITAL-EMERGING-01. RT, ER, MC and GF are with the PETASPIN team and thank the support of PETASPIN association (www.petaspin.com). Work carried out at Tsinghua was supported by the National Natural Science Foundation of China (NSFC) under the distinguished Young Scholar program (Grant No. 12225409), the general program (Grant Nos. 52271181, 51831005, 11861131008), the Beijing Natural Science Foundation (Grant No. Z190009), and the Beijing Advanced Innovation Center for Future Chip (ICFC).

**Keywords**

magnetic multilayer heterostructures**,** interfacial Dzyaloshinskii-Moriya interaction, magnetic skyrmions.